\documentclass[prl,aps,preprintnumbers,floatfix,byrevtex,twocolumn]{revtex4}
\usepackage{epsfig}
\usepackage{graphics}
\newcommand{\be}{\begin{equation}}
\newcommand{\ee}{\end{equation}}
\begin{document}

\preprint{
\vbox{
\hbox{ADP-04-12/T594}
\hbox{DESY 04-074}
}}

\title{Improved chiral properties of FLIC fermions}

\author{S.~Boinepalli$^1$, W.~Kamleh$^1$, D.B.~Leinweber$^1$,
        A.G.~Williams$^1$ and J.M.~Zanotti$^{1,2}$}

\affiliation{$^1$ Special Research Centre for the
        Subatomic Structure of Matter,
        and Department of Physics,
        University of Adelaide, Adelaide SA 5005, Australia}
\affiliation{$^2$ John von Neumann-Institut f\"ur Computing
        NIC, \\
        Deutsches Elektronen-Synchrotron DESY, D-15738 Zeuthen,
        Germany}  

\begin{abstract}
  The chiral properties of the fat-link irrelevant clover (FLIC)
  fermion action are examined.  The improved chiral properties of
  fermion actions incorporating smoothed links are realized in the
  FLIC action where only the irrelevant operators of the fermion
  action are constructed with smoothed links.  In particular, the
  histogram of the additive mass renormalization encountered in
  chiral-symmetry breaking Wilson-type fermion actions is seen to
  narrow upon introducing fat-links in the irrelevant operators.  The
  exceptional configuration problem of quenched QCD is reduced,
  enabling access to the light quark mass regime of $m_{\pi} /
  m_{\rho} \sim 0.35$.  In particular, quenched chiral non-analytic
  behavior is revealed in the light quark mass dependence of the
  $\Delta$-baryon mass.  FLIC fermions offer a promising approach to
  revealing the properties of full QCD at light quark masses.

\end{abstract}

\vspace{3mm}

\maketitle

The frontier of lattice QCD lies at the challenge of directly
simulating the full theory, including the dynamical sea-quark loop
contributions, at light quark masses.
In principle, the problem is solved.  The advent of fermion actions
which provide exact lattice chiral symmetry, in particular the overlap
action\cite{overlap}, allows one to directly simulate at the physical
quark masses realized in QCD.  However the computational cost of such
fermion actions are prohibitively expensive.  Simulations on
physically large volumes with a cutoff sufficiently large to
accommodate the physics of interest are not possible at present.
Indeed significant breakthroughs are required to reduce the
computational cost to a level where leading edge computing resources
can have a significant impact.

In contrast the staggered fermion action is computationally cheap and
is currently having a tremendous impact near the light quark mass
regime \cite{staggered}.  Had nature presented us with four degenerate
light-fermion flavors, the formalism would be ideal.  In this
formalism, the 16-fold degeneracy of fermion flavors encountered in
naively discretising the continuum derivative is
reduced to a 4-fold flavor degeneracy.  In practice, the fermion determinant
describing the contribution of sea-quark loops having their origin in
the gluon field is reduced to a single flavor by taking the
fourth-root.  This operation renders a non-local lattice action,
raising concerns about the continuum limit of the staggered theory
\cite{staggeredConcerns}. 

On the other hand the Wilson fermion action defines a local lattice
action for a single fermion flavor that is computationally efficient
but at the expense of explicitly breaking global chiral symmetry
through the introduction of the lattice Laplacian operator in an
irrelevant, energy dimension-five operator.  This operator solves the
fermion doubling problem by giving the doublers a mass proportional to
the inverse lattice spacing $a$, but at the expense of introducing
large ${\cal O}(a)$ errors.  A systematic approach
\cite{Symanzik:1983dc} to achieving efficient ${\cal O}(a)$
improvement of the lattice fermion action leads to the
Sheikholeslami-Wohlert fermion \cite{Sheikholeslami:1985ij} action, or
``clover action'', so named because it is constructed by adding the
clover term, $(i\, g\, a\, C_{\rm SW}\, / 4)\,
\bar{\psi}(x)\, \sigma_{\mu\nu} F_{\mu\nu}\, \psi(x)$
to the standard Wilson action \cite{Wilson}.
%
The clover coefficient, $C_{\scriptstyle {\rm SW}}$, can be tuned to
remove ${\cal O}(a)$ artifacts to all orders in the gauge coupling
constant $g$.  Nonperturbative methods have been established
\cite{Luscher:1996sc,FATJAMES,FLICscaling} to accurately define
improvement coefficients such as $C_{\scriptstyle {\rm SW}}$ in the
interacting theory.  Traditionally, large renormalizations of the
improvement coefficients have hindered the realization of lattice
action improvement in practice.

While the nonperturbatively improved SW action correctly defines a
lattice action for a single fermion flavor, scales well
\cite{Edwards:1998nh}, and remains computationally efficient, the
breaking of global chiral symmetry makes the approach to the chiral
regime difficult.  Global chiral symmetry breaking introduces an
additive mass renormalization into the Dirac operator that can give
rise to singularities in quark propagators at small quark masses.  The
problem is exacerbated through the use of large lattice volumes; large
values of the strong coupling constant, $g$, providing large lattice
spacings, $a$; or large values of the improvement coefficient,
$C_{\scriptstyle {\rm SW}}$
\cite{Luscher:1996sc,Bardeen:1998gv,DeGrand:1998jq}.  In practice,
this prevents the use of coarse lattices (Wilson: $\beta < 5.7 \sim a
> 0.18$~fm) \cite{Bardeen:1998gv,DeGrand:1998jq}.

The singularities are caused by the shifting of the point where the
renormalized mass is zero, away from the point where the bare mass
parameter $m$ is zero.  The position of the singularity is both gauge
configuration and action dependent \cite{Bardeen:1998gv}.  Bare quark
masses must be shifted by an amount dependent on the gauge action to
restore chiral symmetry.  That is, a ``critical mass'' is introduced
and fine tuned such that the pion mass vanishes when the quark masses
take the critical mass $m=m_{\rm cr}$.  All other quark masses are
measured relative to $m_{\rm cr}$.

The difficulty in approaching the chiral regime becomes apparent once
one realizes the critical mass varies on a gauge-field configuration
by configuration basis.  For fixed input bare mass, the renormalized
mass, relevant to the physical properties of the quark, is different
on every configuration.  Given that most observables encounter rapidly
varying chiral nonanalytic behavior in the chiral regime
\cite{FRRchiPT}, this small variation in the renormalized mass gives
rise to widely varying hadron properties on a configuration by
configuration basis which is ultimately realized as a large
statistical error bar in hadronic observables.
Thus, for the generation of gluon-field configurations incorporating
the effects of dynamical fermion loops, the development of chirally
improved fermion actions which minimize the breadth of the
distribution of the critical mass obtained on a configuration by
configuration basis (while maintaining only nearest neighbor
interactions) is highly desirable.

In the generation of dynamical-fermion gauge configurations the
fermion determinant acts to suppress the probability of creating
configurations giving rise to approximate zeromodes of the Dirac
operator, $D(m)$.  In the quenched approximation, the fermion
determinant is set to a constant and a proliferation of approximate
zeromodes is encountered.  At sufficiently light renormalized quark
mass, it is possible to have the critical mass realized on a
particular configuration near the bare mass, introducing a divergence
in the quark propagator.  Such ``exceptional'' configurations provide
hadronic correlation functions which differ significantly in magnitude
from the average, again introducing large statistical uncertainties
and on some occasions spoiling the ensemble average result.

Fat-link actions \cite{DeGrand:1998jq,DeGrand:1999gp} have been
identified as providing improved chiral properties while maintaining
only nearest neighbor interactions.  Fat links are obtained by
averaging the gauge field links, $U_\mu(x) = \exp[ i g \int_0^a
A_\mu(x + \lambda \widehat \mu) d\lambda]$, with their transverse
neighbors in an iterative process of APE smearing \cite{APE} or HYP
smearing \cite{HYP}.  The incorporation of smoothed links throughout
the fermion action removes short distance interactions and thus
suppresses renormalizations of the critical mass in the interacting
theory.  DeGrand {\it et al.}~\cite{DeGrand:1998jq} noted that the
appearance of spurious zero modes is due to local lattice dislocations
with non-zero topology at the scale of the cutoff.  By smoothing the
gauge field at short distances it was possible to narrow the
distribution of the critical mass by about a factor of three, thus
reducing the problem of exceptional configurations.

However short distance effects are lost in the fat-link theory.  FLIC
fermions circumvent this problem by introducing fat-links only in the
purely irrelevant lattice operators of the clover action
\cite{Sheikholeslami:1985ij}, having energy dimension five or more.
The untouched gauge fields generated via Monte Carlo methods form the
relevant operators associated with the continuum action.  FLIC
fermions provide near continuum results at finite lattice spacing,
while preserving short distance physics.
The FLIC action provides an alternative form of nonperturbative ${\cal
O}(a)$ improvement \cite{FLICscaling} which avoids the fine tuning
problem of improvement coefficients.

The focus of this investigation is to determine the extent to which
the improved chiral properties of fat-link fermion actions are
realized with FLIC fermions.  Previous work
\cite{FATJAMES,Kamleh:2001ff} has shown that the FLIC fermion action
has extremely impressive convergence rates for matrix inversion, which
provides great promise for performing cost effective simulations at
quark masses closer to the physical values.  In the following, we will
illustrate how the distribution of the critical mass $m_{\rm cr}$ is
narrowed sufficiently to allow simulations at light quark masses
providing $m_{\pi} / m_{\rho} \sim 0.35$ while exceptional
configurations, encountered in the quenched approximation, are limited
to $\sim 6\%$.  Access to the light quark mass region is sufficient to
reveal quenched chiral nonanalytic behavior in the quark mass
dependence of the $\Delta$-baryon mass.

Our simulations are based on gauge fields described by the ${\cal
O}(a^2)$-mean-field improved Luscher-Weisz plaquette plus rectangle
gauge action \cite{Luscher:1984xn}. We begin with a study of
the distribution of zero modes on 100 $12^3\times 24$ lattices at
$\beta=4.60$, giving a lattice spacing of $a =
0.116(2)$~fm.  The subsequent calculations of 
baryon masses are performed on a sample of 400 $20^3\times 40$
lattices at $\beta=4.53,\ a = 0.128(2)$~fm. The scale is set via $r_0.$

The mean-field improved FLIC action is \cite{FATJAMES,FLICscaling}
\be
S_{\rm SW}^{\rm FL}
= S_{\rm W}^{\rm FL} - \frac{i\, g\, C_{\rm SW}\, \kappa\,
  r}{2(u_{0}^{\rm FL})^4}\, 
             \bar{\psi}(x)\, \sigma_{\mu\nu}F_{\mu\nu}\, \psi(x)\, ,
\label{FLIC}
\ee
where $F_{\mu\nu}$ is an ${\cal O}(a^4)$-improved lattice definition
\cite{Bilson-Thompson:2002jk} constructed using fat links, $u_{0}^{\rm FL}$ is
the plaquette measure of the mean link calculated with fat links, and
where the mean-field improved Fat-Link Irrelevant Wilson action is
\begin{eqnarray}
S_{\rm W}^{\rm FL}
=  &\sum_x& \bar{\psi}(x)\psi(x) 
+ \kappa \sum_{x,\mu} \bar{\psi}(x)
    \bigg[ \gamma_{\mu}
      \bigg( \frac{U_{\mu}(x)}{u_0} \psi(x+\hat{\mu}) \nonumber \\
&-& \frac{U^{\dagger}_{\mu}(x-\hat{\mu})}{u_0} \psi(x-\hat{\mu})
      \bigg)
- r \bigg(
 \frac{U_{\mu}^{\rm FL}(x)}{u_0^{\rm  FL}} \psi(x+\hat{\mu})\nonumber\\
&+& \frac{U^{{\rm FL}\dagger}_{\mu}(x-\hat{\mu})}{u_0^{\rm FL}}
          \psi(x-\hat{\mu})
      \bigg)
    \bigg]\ .
\label{MFIW}
\end{eqnarray}
with $\kappa = 1/(2m+8r)$. We take the standard value $r=1$.  Our
notation uses the Pauli representation of the Dirac $\gamma$-matrices
\cite{SAKURAI}, where the $\gamma$-matrices are hermitian
and $\sigma_{\mu\nu} = [\gamma_{\mu},\ \gamma_{\nu}]/(2i)$. Fat links
are constructed by performing $n_{\rm APE}$ sweeps of APE smearing,
where in each sweep the weights given to the original link and the
six transverse staples are 0.3 and $(0.7/6)$ respectively.
The FLIC action is closely related to the mean-field improved clover
(MFIC) fermion action in that the latter is described by
Eqs.~(\ref{FLIC}) and (\ref{MFIW}) with all fat-links replaced by
untouched thin links and $F_{\mu\nu}$ defined by the $1 \times 1$-loop
clover definition.

The critical mass for a particular configuration is determined by
examining the spectral flow of the Hermitian FLIC Dirac operator,
$\gamma_5 D(m)$, commencing with quark masses in the physical mass
regime and proceeding towards the negative quark mass regime relevant
to overlap fermions. The physical regime is characterized by the
magnitude of the lowest eigenvalue $\lambda_{\rm min}(m)$ decreasing
as $m$ decreases.  The negative mass regime is characterized by the
presence of eigenvalues crossing zero, due to non-trivial topology.
The larger the topological object, the closer to $m_{\rm cr}$ is the
induced zero crossing \cite{Kusterer:2001vk}.  Hence, we can determine
the regime transition point by calculating $\lambda_{\rm min}(m)$ and
observing it decrease with $m$ until it either crosses zero or turns
away and begins to increase. The critical mass is then defined as the
$m$ at which the transition occurs.

\begin{figure}[tb]
\begin{center}
{\includegraphics[height=0.8\hsize,angle=90]{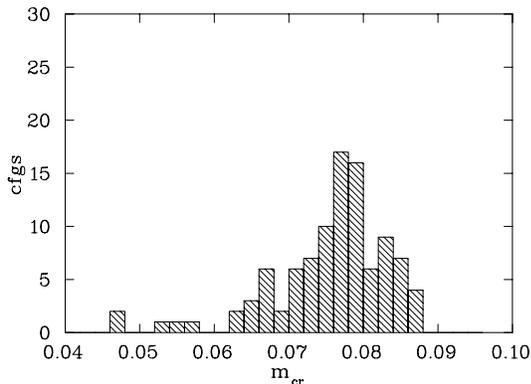}}
\vspace*{-0.4cm}
\caption{Histogram of the critical mass obtained from the spectral
  flow of the mean-field improved clover (MFIC) fermion action on 100
  configurations.}
\label{clovermcr}
\end{center}
\vspace{-0.5cm}
\end{figure}

\begin{figure}[tb]
\begin{center}
{\includegraphics[height=0.8\hsize,angle=90]{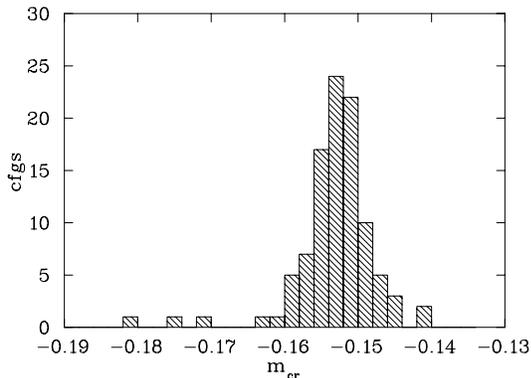}}
\vspace*{-0.4cm}
\caption{Histogram of the critical mass obtained from the spectral
  flow of the fat-link irrelevant clover (FLIC) fermion action ($n_{\rm APE}=4$) on 100
  configurations.}
\label{flicmcr}
\end{center}
\vspace{-0.6cm}
\end{figure}

Figures \ref{clovermcr} and \ref{flicmcr} illustrate histograms of the
critical mass obtained from the MFIC fermion action and FLIC fermion
action respectively.  Each histogram is obtained from the same set of
100 configurations (cfgs). The FLIC fermion distribution of the
critical mass is significantly narrower than that of the MFIC
action, by a factor of two.  Since the nonperturbatively improved
estimate of $C_{\scriptstyle {\rm SW}}$ typically exceeds that of the
mean-field improved estimate, the distribution for the
nonperturbatively improved clover (NPIC) action will be broader than
that of Fig.~\ref{clovermcr}.  Thus FLIC fermions offer a chirally
improved alternative to MFIC and NPIC fermion actions.  FLIC fermions
enjoy a significantly reduced exceptional configuration
problem, and provide reduced fluctuations in hadronic observables as
the chiral limit is approached in full QCD.

To directly examine the chiral properties of the FLIC fermion action,
we determine the low-lying hadron mass spectrum.  Hadron masses are
extracted from the Euclidean-time dependence of the two-point
correlation functions using standard techniques.  Effective masses are
calculated as a function of time and time-fitting intervals and are
selected via standard covariance matrix estimates of the $\chi^2 /
N_{\rm DF}$.  For quark masses lighter than the strange quark mass,
effective mass splittings are calculated and fit using the same
techniques.  By examining mass splittings, excited-state contributions
(less dependent on the quark mass) are suppressed and good $\chi^2 /
N_{\rm DF}$ are found one-to-two time slices earlier.

In searching for exceptional configurations, we follow the
technique used by Della Morte {\it et al.} \cite{TMass}.
In the absence of exceptional configurations, the standard deviation
of an observable is independent of the number of configurations
considered in the average.  Exceptional configurations reveal
themselves by introducing a significant jump in the standard deviation
as the configuration is introduced into the average.  

\begin{figure}[t]
\begin{center}
{\includegraphics[height=0.90\hsize,angle=90]{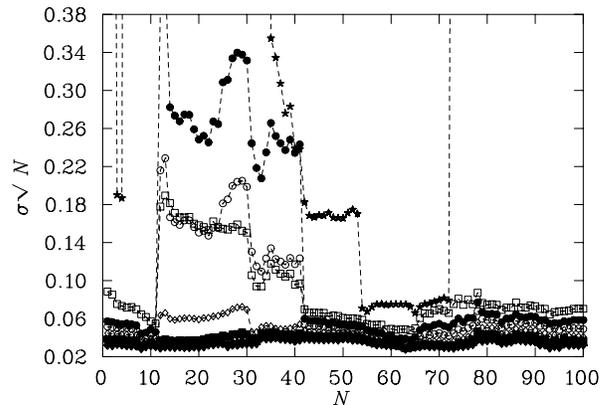}}
\vspace*{-0.4cm}
\caption{ The standard deviation in the error of the $\pi$ mass
  calculated on 30 configurations plotted against the starting
  configuration number for the FLIC-fermion action on a
  $20^3 \times 40$ lattice with $a=0.128(2)$~fm.  See text for details. }
\label{PiErrvsCFG}
\end{center}
\vspace*{-0.7cm}
\end{figure}

Fig.~\ref{PiErrvsCFG} shows the standard deviation of the pion mass
for eight quark masses on subsets of a total of 100 configurations.  A
moving average of 30 configurations is considered, with a cyclic
property enforced from configuration 100 to configuration 1. Configuration 41 
is revealed to be exceptional by noting that between $N=12$ and $N=41$ the 
error blows up for several quark masses and then drops again at
$N=42$ as configuration 41 leaves the moving average. 
The observation of
a large change in the pion mass as 
config.\ 41 in included in the ensemble confirms the exceptional behavior.
Exceptional behavior is also observed for the lightest quark mass for
configurations 2, 13, 30, 34 and 53. Upon removal of these
configurations, a near-constant behavior of the standard deviation is
observed for all quark masses.
Given the relatively coarse lattice spacing, large volume, and the
lightest quark mass providing $m_{\pi} / m_{\rho} \sim 0.35$, an
elimination rate of $\sim 6\%$ is a remarkable result.

\begin{figure}[t]
\begin{center}
{\includegraphics[height=0.9\hsize,angle=90]{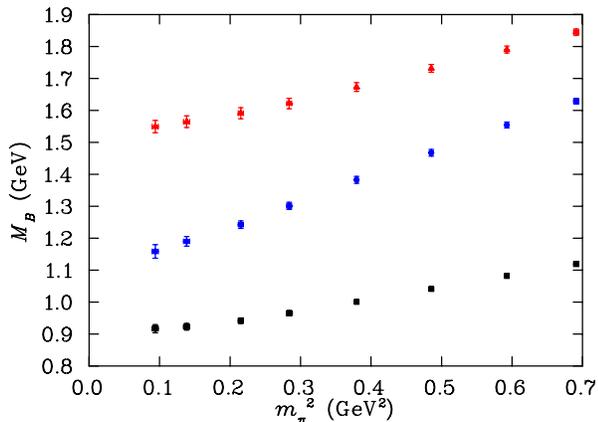}}
\vspace*{-0.4cm}
\caption{$\rho$, $N$ and $\Delta$ masses for the
  FLIC-fermion action ($n_{\rm APE} = 6$) on 400 $20^3 \times 40$
  lattices with $a=0.128(2)$ fm. }
\label{LQM}
\end{center}
\vspace*{-0.9cm}
\end{figure}

Figure~\ref{LQM} shows the $\rho$, $N$ and $\Delta$ masses as a function of
$m_{\pi}^2$ for the FLIC-fermion action.
An upward curvature in the $\Delta$ mass for decreasing quark mass is
observed in the FLIC fermion results.  This behavior, increasing the
quenched $N-\Delta$ mass spitting, was predicted by Young {\em et al}.\
\cite{Young:2002cj} using quenched chiral perturbation theory (Q$\chi$PT)
formulated with a finite-range regulator.  This is the first time that
clear quenched chiral nonanalytic behavior has been observed in a
baryon mass.  Simulations in quenched QCD with Wilson-type fermions on
this lattice spacing for these light quark masses have been previously
unattainable.  Numerical results are summarized in Table
\ref{results}.

\begin{table}
\caption{Values of $\kappa$ and the corresponding $\pi,\, \rho,\,
        {N}$ and $\Delta$ masses (in GeV) for the FLIC fermion action
        on a $20^3 \times 40$ lattice with $a=0.128(2)$ fm.}
\label{results}
\begin{ruledtabular}
\begin{tabular}{ccccc}
    $\kappa$  & $\pi$     & $\rho$     & $N$        & $\Delta$      \\ 
\hline
    0.12780   & 0.831(2)  & 1.119(04)  & 1.629(08)  & 1.845(10) \\
    0.12830   & 0.770(2)  & 1.082(05)  & 1.554(09)  & 1.791(11) \\
    0.12885   & 0.697(2)  & 1.041(07)  & 1.468(11)  & 1.732(12) \\
    0.12940   & 0.616(3)  & 1.001(07)  & 1.383(11)  & 1.673(14) \\
    0.12990   & 0.533(3)  & 0.965(08)  & 1.301(11)  & 1.622(16) \\
    0.13025   & 0.464(4)  & 0.941(09)  & 1.243(12)  & 1.592(17) \\
    0.13060   & 0.372(6)  & 0.923(11)  & 1.190(15)  & 1.565(18) \\
    0.13080   & 0.306(7)  & 0.917(13)  & 1.159(21)  & 1.550(19) \\
\end{tabular}
\end{ruledtabular}
\vspace*{-0.45cm}
\end{table}

In summary, the Fat-Link Irrelevant Clover (FLIC) fermion action is an
efficient lattice fermion operator with excellent scaling properties,
providing near-continuum results at finite lattice spacing.  The
superior chiral properties of the FLIC fermion action enable access to
the light quark-mass regime.  With recent breakthroughs in dynamical
fermion simulations of fat-link fermion actions \cite{Kamleh:2004xk}
FLIC fermions offer a promising approach to revealing the properties
of full QCD at light quark masses.


We thank the Australian Partnership for Advanced
Computing (APAC) and the Australian National Computing Facility for
Lattice Gauge Theory for generous grants of supercomputer time which
have enabled this project.  This work is supported by the Australian
Research Council.
\vspace*{-0.7cm}


\end{document}